\documentclass[prc,aps]{revtex4}

\begin{document}

\title{ Spinodal decomposition of low-density asymmetric nuclear\\
       matter }

\author{V.Baran$^{1,2}$, M. Colonna$^1$, M. Di Toro$^1$ and 
A.B. Larionov$^{1,3}$}

\affiliation{$^1$ Laboratorio Nazionale del Sud, Via S. Sofia 44,
I-95123 Catania, Italy\\ and University of Catania}
\affiliation{$^2$ IFA, Bucharest, Romania}
\affiliation{$^3$ RRC "I.V. Kurchatov Institute", Moscow 123182, Russia}

\begin{abstract}
We investigate the dynamical properties of asymmetric nuclear matter 
at low density. 
The occurrence of new instabilities, that lead the system to a dynamical
fragment formation, is illustrated, discussing in particular the 
charge symmetry
dependence of the structure of the most important unstable modes.
We observe that instabilities are reduced by charge asymmetry, leading 
to larger size and time scales in the fragmentation process. 
Configurations with less asymmetric fragments surrounded by a more asymmetric
gas are favoured.  Interesting variances with respect to a pure thermodynamical
prediction are revealed, that can be checked experimentally.
All these features are deeply related to the structure of the 
symmetry term in 
the nuclear Equation of State ($EOS$) and could be used to extract 
information on the low density part of the $EOS$. 
\end{abstract}
\maketitle
\vspace{1cm}

\hspace{-\parindent}PACS numbers: 21.65.+f, 25.70.Pq, 21.60.Ev

\newpage

\section{ Introduction }

During the last years the equation of state ($EOS$) of nuclear matter 
($NM$) has been studied extensively in the symmetric $N=Z$ case. 
Recent investigations on collisions of radioactive nuclei 
\cite{Pak,Bali,Zyr,Sob} and on formation and structure of 
neutron stars \cite{Prakash} have driven the attention to the 
properties of strongly asymmetric, $N > Z$,
nuclear matter. Hence it appears of relevant interest to 
investigate equilibrium
and non-equilibrium features of asymmetric $NM$ and their connection 
to the used extension of the $EOS$.

In this work we study the influence of charge asymmetry on the spinodal 
decomposition ("$SD$") of nuclear matter at subsaturation density. 
The "$SD$" is the growth of small density perturbations, that leads 
to liquid-gas phase separation, in initially uniform matter located 
in the low density instability region of the $EOS$ phase diagram. 
In symmetric nuclear matter, the "$SD$" was studied extensively in 
Ref.s \cite{Peth87,Peth88,Peth93,Mar93,Mar94,LM94,Mar941,CNP95}. 

To our knowledge, the first publication on the liquid-gas phase separation 
in asymmetric (neutron star) matter is Ref. \cite{Baym71}, where 
the coexistence of asymmetric nuclei with a pure neutron gas was predicted,
at zero temperature. 
Recently, the liquid-gas phase transition in asymmetric nuclear 
matter at finite temperature was studied in Ref.s\cite{Jens,MS,Bali1,Ray}, 
always in an equilibrium thermodynamical approach. For instance, 
it was demonstrated in Ref.\cite{Jens}, on the basis of the Quantum Statistical Model, 
that light clusters emitted from neutron-rich systems have, in average, 
a larger relative neutron excess than the initial source. 

Our discussion is more focussed on non-equilibrium properties of
asymmetric nuclear matter and, in particular, on the possibility to observe
collective 
dynamical formation of clusters, with specific mass and charge contents,
{\it on short time scales}. We expect this mechanism to be important
in fast expanding systems, as in collisions of beta-unstable nuclei,
or as a first step towards preferential equilibrium structures.
The latter point, on possible different time scales in the clustering
process, will be also addressed in this article. 

We consider the unstable growth of density perturbations on the basis
of two Vlasov equations, for neutron and proton liquids, coupled through
the mean field. The influence of the initial asymmetry on the wave length and
growth time of the unstable modes is studied. 
In neutron-rich NM
the formation of larger fragments is favoured and the fragment 
formation process is delayed in comparison to the case of symmetric NM. The 
restoration of the isotopic symmetry inside the formed heavy fragments, predicted 
earlier in Refs. \cite{Bali1,Mar97}, is clearly observed in the "$SD$"
collective dynamical mechanism. Moreover, new non-equilibrium features 
of the fast clustering processes in asymmetric nuclear systems, 
that cannot be explained within a thermodynamical approach, are discussed. 
For instance, for neutron 
excess systems the proton fraction in the gas phase is expected to be larger 
when compared to
statistical predictions \cite{Jens}. This effect can be
related to the {\it "freeze-out"} time, preventing the chemical equilibration 
in the dynamics of heavy ion collisions. 


The structure of the article is as follows. In Sect. II a description
of the mean-field kinetic approach to the coupled neutron and proton 
liquids is given. We apply the linear response analysis 
\cite{Peth87,Peth88,Peth93,Mar93,Mar94,LM94,Mar941,CNP95} for the onset of 
the "$SD$" instabilities. In parallel, we consider numerical 
simulations \cite{Vir} which reproduce all stages of the "$SD$" evolution, 
including non-linear effects. Results from analitical and numerical 
solutions of the coupled Vlasov equations are presented in Sects. III, IV
respectively. Summary and conclusions are given in Sect. V. 

\section{ Theoretical approach }

We start from a mean field description of nuclear dynamics based
on two Vlasov equations, for neutron
and proton liquids \cite{Haensel,Vir,Mar97},
coupled through a self-consistent nuclear field :
\begin{equation}
{\partial f_q({\bf r},{\bf p},t) \over \partial t} 
+ {{\bf p} \over m}{\partial f_q \over \partial {\bf r}}  
- {\partial U_q({\bf r},t) \over \partial {\bf r}}
{\partial f_q \over \partial {\bf p} } = 0~.                  \label{ve}
\end{equation}
Here the subscript $q$ stands for $n$ (neutrons) or $p$ (protons) and
$f_q({\bf r},{\bf p},t)$ is the phase-space distribution function (d.f.).
For simplicity, in Eq.(1) we neglect effective mass corrections and the 
difference between neutron and proton masses, putting $m^*_n = m^*_p = 
m = 938~MeV$. 
Indeed in the low density region studied here we do not expect to have 
large effective mass corrections. 
Finally, $U_q({\bf r},t)$ is the self-consistent mean field potential 
in a Skyrme-like form \cite{Mar97}~:
\begin{equation}
U_q = {\delta{\cal H}_{pot} \over \delta\rho_q} =
  A\left({\rho \over \rho_0}\right) 
+ B\left({\rho \over \rho_0}\right)^{\alpha +1} 
+ C\left({\rho' \over \rho_0}\right)\tau_q
+ {1 \over 2} {d C(\rho) \over d \rho}
{\rho'^2 \over \rho_0} - D\triangle\rho
                       + D'\tau_q\triangle\rho'~,         \label{Uq}
\end{equation}
where
\begin{equation}
{\cal H}_{pot}(\rho_n,\rho_p)=
{A \over 2}{\rho^2 \over \rho_0} +
{B \over \alpha + 2}{\rho^{\alpha + 2} \over \rho_0^{\alpha + 1}} +
               {C(\rho) \over 2} {\rho'^2 \over \rho_0}
+ {D \over 2}(\nabla\rho)^2
- {D'\over 2}(\nabla\rho')^2~                           \label{hpot}
\end{equation}
is the potential energy density;
 $\rho = \rho_n + \rho_p$ and 
$\rho' = \rho_n - \rho_p$ are respectively the total (isoscalar) and the 
relative (isovector) density; $\rho_0 = 0.16~\mbox{fm}^{-3}$ is the 
nuclear saturation density; $\tau_q$ = +1 ($q=n$), -1 ($q=p$).

The values of the parameters $A=-356.8$ MeV, $B=303.9$ MeV, $\alpha=1/6$
and $D=130$ MeV$\cdot$fm$^5$ are adjusted to reproduce the saturation 
properties of symmetric nuclear matter (binding energy
$\epsilon_b=15.7$ MeV/nucleon at $\rho=\rho_0$, zero pressure at
$\rho=\rho_0$, compressibility modulus $K=201$ MeV) and the surface
energy coefficient in the Weizs\"acker mass formula $a_{surf}=18.6$ MeV.
We put $D'= 40$ MeV$\cdot$fm$^5 \sim D/3$ according to Ref.\cite{Baym71},
that is also close to the value $D' = 34$ MeV$\cdot$fm$^5$ given by 
the SKM$^*$ interaction \cite{Kri80}. Thus, the term $\propto D'$ in the 
potential energy density (\ref{hpot}) favours the growth of isovector 
density fluctuations \cite{Baym71}. 

The potential symmetry energy coefficient 
is equal to:
$C(\rho) = C_1 - C_2(\rho/\rho_0)^\alpha$, with $C_1 = 124.9$ MeV
and $C_2 = 93.5$ MeV, where the density dependence corresponds
to the Skyrme energy density functional of a general kind
(c.f. Ref.s\cite{Mar97,Kri80,Sum}). At saturation density the potential
symmetry energy coefficient $C(\rho_0)=31.4$ MeV satisfies 
the condition \cite{Migdal}:
\[
a_{sym} = {\epsilon_F \over 3} + {C(\rho_0) \over 2}~,
\]
where $a_{sym}=28$ MeV is the symmetry energy coefficient in the
Weizs\"acker mass formula, $\epsilon_F^{eq}=36.9$ MeV is the Fermi
energy for the symmetric system at $\rho=\rho_0$.

As a first step, we apply the linear response analysis 
to Vlasov Eqs. (\ref{ve}).
For a small amplitude perturbation of 
the d.f.,
periodic in time,  $\delta f_q({\bf r},{\bf p},t) \sim \exp(-i\omega t)$ 
we can linearize Eqs. (\ref{ve})~:
\begin{equation}
-i\omega\delta f_q 
+ {{\bf p} \over m}{\partial\delta f_q \over \partial {\bf r}}
- {\partial U_q^{(0)} \over \partial {\bf r}}
  {\partial\delta f_q \over \partial {\bf p}}
- {\partial\delta U_q \over \partial {\bf r}}
  {\partial f_q^{(0)} \over \partial {\bf p}} = 0~,   \label{lve}
\end{equation}
where the superscript $(0)$ labels stationary values and 
$\delta U_q$ is the dynamical component of the mean field potential.
The unperturbed d.f. $f_q^{(0)}$ is in general a Fermi distribution at
finite temperature~:
\begin{equation}
f_q^{(0)}(\epsilon_p^q) = 
{1 \over \exp{(\epsilon_p^q-\mu_q)/T} + 1}~,          \label{Fd}
\end{equation}
where $\epsilon_p^q = p^2/(2m) + U_q^{(0)}$ and $\mu_q$ are
respectively energy and chemical potential of the nucleons of 
type $q$. In the present work, we neglect finite size effects
and consider space-uniform unperturbed d.f. Thus, 
$\nabla_r U_q^{(0)} = 0$ in Eq. (\ref{lve}) and we consider
plane-wave solutions 
$\delta f_q \propto \exp(-i\omega t + i{\bf k r})$. 
Following a standard Landau procedure \cite{Peth88,Mar97}, one can derive 
from Eqs. (\ref{lve}) the following system of two equations for 
neutron and proton density perturbations~:
\begin{eqnarray}
& & [1 + F_0^{nn}\chi_n]\delta\rho_n 
  + [F_0^{np}\chi_n]\delta\rho_p      =   0~,       \label{eq1} \\
& & [F_0^{pn}\chi_p]\delta\rho_n
  + [1 + F_0^{pp}\chi_p]\delta\rho_p  =   0~,       \label{eq2}
\end{eqnarray}
where
\begin{equation}
\chi_q(\omega,{\bf k}) = {1 \over N_q(T)} 
\int\,{2~d{\bf p} \over (2\pi\hbar)^3} 
      {{\bf kv} \over \omega + i0 - {\bf kv}}
      {\partial f_q^{(0)} \over \partial\epsilon_p^q}~,  \label{Lfun}
\end{equation}
is the long-wave limit of the Lindhard function \cite{Peth88}; 
${\bf v}={\bf p}/m$;
\begin{equation}
N_q(T) = -\int\,{2~d{\bf p} \over (2\pi\hbar)^3} 
        {\partial f_q^{(0)} \over \partial\epsilon_p^q}
\simeq N_q(0)\left[
1 - {\pi^2 \over 12}\left(T \over \epsilon_{F,q}\right)^2
             \right]~,                                   \label{ldens}
\end{equation}
is the thermally averaged level density ($N_q(0) = mp_{F,q}/(\pi^2\hbar^3)$,
$\epsilon_{F,q}=p_{F,q}^2/(2m)$, $p_{F,q}=\hbar(3\pi^2\rho_q)^{1/3}$)
and, finally
\begin{equation}
F_0^{q_1q_2}(k) = N_{q_1}(T){\delta U_{q_1} \over \delta \rho_{q_2}}
~,~~~~~q_1=n,p,~~~~q_2=n,p                                \label{Lpar}
\end{equation}
are the usual zero-order Landau parameters, where the $k$-dependence is 
caused by the presence of space derivatives in the potentials (see 
Eq.(\ref{Uq})).
For the particular choice of potentials given by Eq.(\ref{Uq}), 
the Landau parameters are 
expressed as \cite{Comm0}~:
\begin{eqnarray}
& & F_0^{q_1q_2}(k) = N_{q_1}(T)\left[ {A \over \rho_0} 
+ (\alpha + 1)B{\rho^\alpha \over \rho_0^{\alpha + 1}}
+          Dk^2  
+ ( {C \over \rho_0} - D'k^2 )\tau_{q_1}\tau_{q_2}\right. \nonumber \\
& & + \left. {d C \over d \rho}{\rho' \over \rho_0}
           (\tau_{q_1} + \tau_{q_2})
+          {d^2 C \over d \rho^2}{\rho'^2 \over 2\rho_0}
      \right]~.                                             \label{Lpar1}
\end{eqnarray}
The dispersion relation connecting eigenfrequencies to wave vectors can be 
obtained by taking the determinant of the system (\ref{eq1}), (\ref{eq2}) 
equal to zero~:
\begin{equation}
(1 + F_0^{nn}\chi_n)(1 + F_0^{pp}\chi_p) 
- F_0^{np}F_0^{pn}\chi_n\chi_p = 0~.                    \label{drel}
\end{equation}

Since we are interested in the unstable growth, we put in Eq.(\ref{Lfun})
$\omega = i\gamma$, where $\gamma > 0$ is the growth rate of the 
instability. Then, in the particular case of $T=0$ one can 
obtain the following 
simple expression for the Lindhard function (\ref{Lfun}) 
(see Ref.\cite{Peth88})~:
\begin{equation}
\chi_q(s_q) = 1 - s_q\arctan(1/s_q)~,                   \label{Lfun1}
\end{equation}
where $s_q = \gamma/(kv_{F,q})$. We stress that without considering the 
wave number
dependence in the Landau parameters (\ref{Lpar1}) one has 
$\gamma\propto k$ and, hence, an unphysical growth of the short wave length
perturbations is favoured \cite{Peth87,LM94,Mar941}.

Stability conditions of asymmetric nuclear matter against density
fluctuations are derived in the Appendix. The system becomes unstable
if at least one of conditions (\ref{c1}), (\ref{c2}) is violated.
 
The linear technique described above is applicable only in the
regime of small-amplitude perturbations, and some relevant
results will be shown in the next section. Then as a second step,
see Sect. IV,
we consider the more powerful numerical solution of the Vlasov Eqs. (\ref{ve}),
which is based on the test-particle approach \cite{Gre87,BGM}.
The detailed description of the numerical method is given in Refs.
\cite{Vir,Vir1}. In this way, we can also have effects from nonlinear terms
and particle collisions. The latter contribution is indeed always present
in the dynamical response of heated nuclear matter. However, the main results
shown in the next section from a pure mean field approach should not
be much affected, since a very dilute system is considered.  This will
be also confirmed from comparisons with numerical simulations, where
the collision integral is included (see Sect. IV).

\section{ Unstable solutions of extended Landau dispersion relations}

We discuss, first, the results of the linear response theory. Fig. 1 shows the
instability region (under curves), as given by the inequality:
\begin{equation}
(1 + F_0^{nn})(1 + F_0^{pp}) - F_0^{np}F_0^{pn} < 0~,     \label{instcon}
\end{equation}
with the Landau parameters taken at $k= 0$
(see the stability condition (\ref{c2}) and text below in the Appendix) 
in the $\rho-T$ plane for
different asymmetries $I= (N-Z)/A$ (a) and in the $\rho-I$ plane 
for different 
temperatures (b).
The asymmetry leads to shrinking of the spinodal region, reducing
both critical temperature and density (Fig. 1a), in agreement
with the results of ref.\cite{MS}. This is indeed a quite general effect
due to the attractive neutron-proton effective interaction and repulsive
neutron-neutron and proton-proton ones \cite{Haensel}:
a repulsive symmetry term is softening the $EOS$ for asymmetric $NM$
reducing then the low density instability region.

An increasing temperature also
reduces the unstable region in the $\rho-I$ plane (Fig.1b). 

We have solved the dispersion relation eq.(\ref{drel}) looking at 
isoscalar growing modes
$(\delta\rho_p/\delta\rho_n > 0)$, considering various choices of the 
initial density,
temperature and asymmetry of nuclear matter.
Fig. 2 reports the growth rate $\Gamma~=~{\rm Im}~\omega(k)$
 as a function of the wave vector $k$. 
The growth rate has a maximum $\Gamma_0=0.01\div0.03$ c/fm corresponding 
to a wave vector value around $k_0=0.5\div1~\mbox{fm}^{-1}$ and 
becomes equal to zero
at $k\simeq1.5k_0$, due to $k$-dependence of the Landau parameters, as 
discussed above. 
One can see also that instabilities are reduced when increasing the
temperature. This effect is present also in the symmetric N = Z case 
\cite{Mar941}.

At larger initial asymmetry the development 
of the "$SD$" is slower. One should expect also an increasing of 
the size of the produced 
fragments due to the "$SD$" features in asymmetric systems. The effect of 
the asymmetry on the growth time $t_0=1/\Gamma_0$ and on the wave length 
$\lambda_0=2\pi/k_0$ of the most unstable mode is shown in Fig. 3. 
It is quite clear that the asymmetry dependence of both variables
$\Gamma_0$ and $\lambda_0$ is more pronounced at higher temperature,
when the system is closer to the boundary of the spinodal region. 

A better understanding of the "$SD$" in a two-component system can be achieved
by studying the chemical composition of the growing mode. On Fig. 4 we show
the asymmetry of the perturbation 
$I_{pt}=(\delta\rho_n - \delta\rho_p)/(\delta\rho_n + \delta\rho_p)$
as a function of the asymmetry of the initially uniform system
$I=(\rho_n^{(0)} - \rho_p^{(0)})/(\rho_n^{(0)} + \rho_p^{(0)})$.
Without any chemical processes, we should expect $I_{pt}=I$.
However, we obtain $I_{pt}\simeq0.5~I$. This means that a growing mode
produces more symmetric high-density regions (liquid phase) and
less symmetric low-density regions (gas phase). Hence, during the "$SD$",
a collective diffusion of protons from low-density regions to high-density
regions takes place. 

We see from Fig. 4 that the chemical effect becomes stronger with increasing
density. This can be explained by the increasing behaviour of the symmetry 
energy per nucleon with density, in the density region considered here. 
The effect of increasing the temperature goes in the opposite direction, 
reducing the chemical effect.

{\it The conclusion is that the fast "$SD$" mechanism in a neutron-rich 
matter will dynamically form more symmetric fragments surrounded by a less 
symmetric gas.} 
Some recent experimental observations from fragmentation reactions
with neutron rich nuclei seem to be in agreement with this result on
the fragment isotopic content : nearly symmetric Intermediate Mass
Fragments ($IMF$) have been detected surrounded by very neutron-rich
light ions \cite{Yen97}.
 
\section{Numerical results: heated nuclear matter in a box}

The previous analytical study is restricted to the onset of the "$SD$",
in a linearized approach.
Then numerical calculations were performed in order to study all
stages of the fragment formation process.
In the numerical approach we consider nuclear matter in a cubic box of 
size $L$ imposing periodic boundary conditions. 
We follow a phase-space test particle method to solve the Landau-Vlasov 
dynamics,
using gaussian wave packets \cite{Gre87,BGM,Vir1}. 
 The dynamics of nucleon-nucleon collisions is included by solving 
the Boltzmann-Nordheim collision integral using a Monte-Carlo method
\cite{BGM}.

We choose the width 
of the gaussians in order
to correctly reproduce the surface energy value in finite systems. 
In this way a cut-off appears in the short wavelength unstable modes,
preventing the formation of too small, unphysical,  clusters \cite{Mar941}. 
In order to have a correct mean field treatment also on the edges of the box, we
have used a
self-consistent {\it stuffing} method, filling a layer of $6fm$ thickness all
around the box with test particles having symmetric positions with respect to
those on the opposite side in the box. 
  
  The calculations were performed using 80 gaussians per nucleon and the
number of nucleons inside the box was fixed in order to reach the initial 
uniform density value. An initial temperature is introduced by distributing the
test particle momenta according to 
Fermi d.f. (Eq. (5)).
We have checked that we reproduce at equilibrium 
the right "EOS" corresponding to the used effective forces (see Sect. II). 

 We have followed the space-time
evolution of test-particles in the box with side $L=~24fm$ for three values of 
the initial
asymmetry $I=0,~0.25$ and $0.5$, at initial density 
$\rho^{(0)}=0.4\rho_0$ and temperature $T=5$ MeV. The initial density 
perturbation was created automatically due to the random choice of 
test-particle positions. 
Results for the initial asymmetries $I=0$ and
$I=0.5$, are reported in Fig. 5, (a) and (b) respectively.  
Figure 5 shows density distributions in the 
plane $z=0$, which contains the center of the box, at three time steps
$t=0,~100$ and $200$ fm/c, corresponding respectively to initial 
conditions, intermediate and final stages of the "$SD$". Clearly, the 
growth of the small initial density perturbations takes place. 

We have compared the dynamical evolution, as given by the test particle method, 
with the analytical predictions of Sect. III (Figs. 2-4). 
To do this, two variables were 
constructed: the total density variance (see \cite{Mar93})
\begin{equation}
         \sigma = < (\rho - \rho^{(0)})^2 >_{all}        \label{totvar}
\end{equation}
and the correlation function between proton and neutron density perturbations,
normalized to the neutron density variance,
\begin{equation}
         R_{pn} = {< (\rho_p - \rho_p^{(0)}) (\rho_n - \rho_n^{(0)})
         >_{all}   \over < (\rho_n - \rho_n^{(0)})^2 >_n }~.
                                                         \label{Rpn}
\end{equation}
In Eqs. (\ref{totvar}),(\ref{Rpn}) $<...>_{all}$ denotes the average
over all test particles, while $<...>_n$ denotes the average over
neutrons only.  The densities $\rho$, $\rho_n$ and $\rho_p$ 
were calculated in the position of the test particle considered by taking 
contributions from gaussians of all test particles. For a 
dominant plane-wave perturbation we have:
\begin{eqnarray}
\sigma &\propto& \exp(2\Gamma t),                       \nonumber \\
R_{pn} &=& {\delta\rho_p \over \delta\rho_n}~.         \nonumber
\end{eqnarray}
Fig. 6 shows the evolution of 
$\sigma$ (a) and of the
(test-particle) perturbation asymmetry $I_{pt}=(1-R_{pn})/(1+R_{pn})$
 (b),
for the same initial conditions discussed above, i.e. $T=5~MeV$,
$\rho^{(0)}=0.4\rho_0$ and asymmetries $I=0.0,~0.25,~0.5$.

A general feature is the
 clear linear increase of $ln(\sigma)$ in the time
interval
$50 < t < 150$ fm/c. During the first $50~fm/c$ the system is
quickly "self-organizing" selecting the most unstable normal
mode. Afterwards the variance (Eq. (\ref{totvar})) increases exponentially
with a time scale given by $\Gamma={\rm Im}~\omega(k)$.
In correspondence (see Fig.6b), the perturbation asymmetry $I_{pt}$
reveals also a quick saturation at 
$t \sim 50$ fm/c. At earlier times the proton and
neutron density perturbations are not correlated, but at 
$t > 50$ fm/c the correlation of plane-wave type 
$(\delta\rho_p/\delta\rho_n=\mbox{const} > 0)$ develops.

We notice that the time scales necessary to reach the asymmetry value
characteristic of the most important growing modes, that are quite 
short in our calculations, generally depend on the structure of the 
initial noise put in the neutron and proton densities. In our calculations
all modes are nearly equally excited. This causes the quick appearance of 
the features associated with the dominant mode. In agreement with analytical 
calculations, the instability grows slower in the case of larger asymmetry.

For an initial asymmetry $I=0.5$,
the extracted values of growth time $\Gamma \simeq 0.01$ c/fm and perturbation
asymmetry $I_{pt} \simeq 0.24$ (see Fig. 6), and of wave length 
$\lambda \simeq 12$ fm (from the distance between the density distribution 
maxima in Fig. 5b),
are in good agreement with the analytical 
results presented in Figs. 3,4.

The "$SD$" leads to a fast formation of the liquid 
(high density) and gaseous (low density) phases in the matter. 
Indeed this dynamical mechanism of clustering will roughly end
when the variance (Eq. (\ref{totvar})) saturates \cite{cdg94}, i.e.
around $250~fm/c$ in the asymmetric cases (see Fig. 6a).

We will discuss in the following the "chemistry" of the liquid
phase formation. In Fig. 7 we report the time evolution of
neutron (thick histogram in Fig. 7a) and proton (thin histogram in Fig. 7a)
abundances and asymmetry (Fig. 7b) in various density bins.
The dashed line respectively shows the initial uniform density
value $\rho \simeq 0.4\rho_0$ (Fig. 7a) and the initial asymmetry
$I=0.5$ (Fig. 7b). The drive to higher density regions is clearly 
different for neutrons and protons: at the end of the dynamical 
clustering mechanism we have very different asymmetries in the 
liquid and gas phases (see the panel at $250fm/c$ in Fig. 7b).

It was shown in Refs. \cite{Baym71,MS,Bali1}, on the basis of 
thermodynamics, that the two phases should have different asymmetries,
namely, $I_{gas} > I_{liquid}$, and actually a pure neutron gas
was predicted at zero temperature if the initial global asymmetry 
is large enough ($I>0.4$) \cite{Baym71}. Here we are studying this 
chemical effect in a non-equilibrium clustering process, on very short 
time scales. The interest is in the observation of new features, not 
expected in a thermodynamical picture.

In our numerical model, we divide the system 
into liquid-like and gas-like phases as follows. The $i$-th test particle 
belongs to the liquid- (gas-) like phase if $\rho_i > (<)~\rho(t=0)$, 
where $\rho_i$ is the total density in the position of the $i$-th 
test particle,
$\rho(t=0)$ is the initial density \cite{Comm}. 

Fig. 8 presents the time evolution of the asymmetry in liquid and in gas.
Strong fluctuations take place at the beginning of the evolution. 
However, at $t > 50$ fm/c, results are quite stable.
In the symmetric case
($I=0$), both liquid and gas keep the symmetry in the course of the time 
evolution. 
{\it 
However for non-zero initial asymmetry liquid becomes more 
symmetric and gas less symmetric (neutron-rich) as time goes.}
The asymmetry in liquid and gas phases saturates rather early, at 
$t = 200\div300$ fm/c, just on the time scale of dynamical clusterization. 

This suggests an evidence of a two-stage fragment formation process. 
On the first stage, the fast "$SD$" takes place ($t < 250$ fm/c,
see Fig. 6a). At the end of this stage, due to reduced asymmetry of the 
perturbation, see Fig. 4, liquid acquires a lower asymmetry and 
gas acquires a higher 
one. On the second stage ($t > 250$ fm/c) we have
a statistical nucleation process: protons and neutrons 
diffuse very slowly from gas to liquid, see Fig. 9a,
 with approximately equal rates (Fig. 9b). 
Since the gas is already highly asymmetric and the 
liquid is more close to symmetry, 
this diffusion process leads to a further growth of asymmetry in the gas and 
does not change the asymmetry of the liquid.
Eventually, on long time scales, we can expect to reach the
thermodynamical limit of Refs. \cite{Baym71,MS,Bali1}, i.e.
an almost pure neutron gas.

From our dynamical analysis we can conclude that the second
slow mechanism for liquid formation is certainly present
in a confined system and will be stopped in an expanding
case. We deduce that the isotopic contents of the gas phase
could give a measure of the "{\it freeze-out}" time: a very neutron-rich
gas will correspond to a slowly expanding nuclear system.

This effect is quantitatively shown in Fig. 10 where we present
asymmetries of the two phases at different time stages,
as given by the test-particle simulations. 
We choose two "{\it freeze-out}" times, one corresponding to the
intermediate stage of the "$SD$" ($t=150$ fm/c) and the second inside
the slow diffusion process ($t=400$ fm/c). The first choice of freeze-out 
time is close to the time scale of the fast multifragment breakup in 
intermediate energy heavy ion collisions (c.f. \cite{Mar941,cdg94,BLM}). 

The asymmetry of the liquid is, practically, not dependend on the choice of 
"{\it freeze-out}" time. On the contrary, the asymmetry of the gas and the number 
of gas particles are very sensitive to this choice.

\section{ Summary and conclusions }

An investigation of fragment formation through Spinodal Decomposition
in low-density asymmetric nuclear matter has been performed.
Important information on the early evolution of the unstable modes is obtained
performing an 
analytical linear analysis to the  
Vlasov equations, for neutron and proton 
liquids, coupled through the mean field. 
Then a  numerical study of nuclear systems with periodic 
boundary conditions has been applied to describe all stages of the fragment 
formation process. 

We have shown that charge asymmetry changes time and space scales 
of the fragment formation process. This effect could be observed 
experimentally.

With respect to the fragment
isotopic distribution, we predict that
the "$SD$" in asymmetric systems is accompanied by a collective
diffusion of protons from low to high density regions. Thus the  
produced heavy fragments will be more symmetric than the initial uniform 
system. This qualitatively 
corroborates earlier thermodynamical studies \cite{Bali1}. 
However the chemical composition of fragments is established just 
after the finishing of the fast "$SD$" stage. Hence it is essentially defined
by the non-equilibrium mean field dynamics and a thermodynamical approach 
gives only a rough schematic description of this process, with a slow
nucleation component likely absent in fragmentation reactions
with radioactive beams. In general in a dynamical multifragmentation process 
with neutron rich systems we expect a proton fraction in the gas phase,
i.e. among emitted nucleons and light clusters, larger than the 
thermodynamical prediction. 


Finally we think that different choices of effective interaction,
(eg. with different density dependence in the symmetry energy coefficient 
$C(\rho)$) 
will support the qualitative conclusions of our work, since all realistic effective 
interactions give a similar behaviour for the potential symmetry energy 
per nucleon 
in the region of subnuclear densities \cite{Bombaci}. 
However the size of the instability region and the isotopic structure
of the most unstable collective modes will be certainly dependent
on the used effective force. Therefore, considering
fragmentation reactions with radioactive beams, it would be 
possible to extract
important information on the low density part of the $EOS$
in asymmetric nuclear matter, of large interest for the
understanding of life and structure of neutron stars.

\vskip 1.0cm

{\bf Acknowledgements}

We gratefully acknowledge intense and stimulating discussions within
the $ISO-DYN$ working group, international working group on
isospin effects on nuclear dynamics. Two of us, V.B. and A.B.L., 
acknowledge kind hospitality and financial support of LNS-INFN. 

This work was supported in part by the Commission of the European Community,
under Contract No. ERBFMB I-CT-960654.

\section*{Appendix: Stability conditions}

Here we derive the stability conditions of asymmetric nuclear matter against
density perturbations. Let us write down the variation
of the free energy density ${\cal F} = \epsilon - T\sigma$, where
$\epsilon$ and $\sigma$ are respectively energy and entropy densities,
over proton and neutron densities, up to the second order \cite{Baym71}
keeping constant temperature and volume~:
\begin{equation}
\delta{\cal F} = \mu_n\delta\rho_n + \mu_p\delta\rho_p 
+ {\partial\mu_n \over \partial\rho_n}{\delta\rho_n^2 \over 2}
+ {\partial\mu_p \over \partial\rho_p}{\delta\rho_p^2 \over 2}
+ {\partial\mu_p \over \partial\rho_n}\delta\rho_n\delta\rho_p~.
                                                            \label{deltaF}
\end{equation}
First two terms disappear after integration over the volume, since the total
neutron and proton numbers are conserved~:
\begin{equation}
\int\,d^3 r \delta{\cal F} 
= {1 \over 2}\int\,d^3 r \left[ 
{\partial\mu_n \over \partial\rho_n}\delta\rho_n^2 
+ {\partial\mu_p \over \partial\rho_p}\delta\rho_p^2
+ 2 {\partial\mu_p \over \partial\rho_n}\delta\rho_n\delta\rho_p
\right]~.                                 
                                                     \label{intF}
\end{equation}
In order the system to be stable against density fluctuations, 
the quadratic form in square brackets of eq. (\ref{intF}) must be 
positive defined. Thus, we obtain two stability conditions~:
\begin{eqnarray}
& & {\partial\mu_n \over \partial\rho_n} > 0~~~~ \mbox{or}~~~~  
{\partial\mu_p \over \partial\rho_p} > 0~,           \label{cond1} \\
& & {\partial\mu_n \over \partial\rho_n}
{\partial\mu_p \over \partial\rho_p} 
- \Big({\partial\mu_p \over \partial\rho_n}\Big)^2 > 0         \label{cond2}~.
\end{eqnarray}
The system is unstable if at least one condition 
is violated. 
We notice that 
it is enough to satisfy only one inequality in (19), since the other one will
be satisfied automatically, if condition (20) is fulfilled.
In an equivalent way, the stability conditions
(\ref{cond1}) and (\ref{cond2}) can be expressed in terms of the Landau
parameters as follows~:
\begin{eqnarray}
& & 1 + F_0^{nn} > 0~~~~\mbox{or}~~~~1 + F_0^{pp} > 0~,   \label{c1} \\
& & (1 + F_0^{nn})(1 + F_0^{pp}) - F_0^{np}F_0^{pn} > 0~, \label{c2}
\end{eqnarray}          
where we have used the relation
\begin{equation}
N_q(T){\partial\mu_q \over \partial\rho_{q'}}
= \delta_{qq'} + F_0^{qq'},~~~~~~~q=n,p~~~~~~q'=n,p    \label{muF0}
\end{equation}
with $\delta_{qq'}=1$ if $q=q'$ and $\delta_{qq'}=0$ if $q \neq q'$.
The relation (\ref{muF0}) is obtained by considering the variation over
$\delta\rho_{q'}$ of both sides of the equation: 
\begin{equation}
\rho_q = \int\,{2~ d{\bf p} \over (2\pi\hbar)^3}
f_q^{(0)}(\epsilon_p^q)~.                              \label{rhoq}
\end{equation}
In the case of the interaction used here (see Eq. (\ref{Lpar1})), the condition 
(\ref{c1}) is always satisfied. 

The stability condition (\ref{cond2})
was derived first in Ref. \cite{Baym71}. Then it was shown in
Ref. \cite{MS} that this stability condition 
can be
rewritten equivalently as:
\begin{equation}
\left({\partial P \over \partial \rho}\right)_{T,y}
\left({\partial\mu_p \over \partial y}\right)_{T,P} > 0~,
                                                         \label{cc2}
\end{equation} 
where $y=\rho_p/\rho$ is the proton fraction, 
$P = \rho\left({\partial{\cal F} \over \partial\rho}\right)_{T,y}
- {\cal F}~$ is the pressure.  

The boundary of the instability region
can be obtained just putting $\omega=0$ in the
Lindhard function (\ref{Lfun}). Since $\chi_q(\omega=0,k)=1$, we have
from dispersion relation (\ref{drel})~:   
\begin{equation}
(1 + F_0^{nn})(1 + F_0^{pp}) - F_0^{np}F_0^{pn} = 0~.     \label{bc2}
\end{equation}
In the case of symmetric nuclear matter 
$(F_0^{nn}=F_0^{pp},~F_0^{np}=F_0^{pn})$ the stability conditions
(\ref{c1}),(\ref{c2}) are equivalent to the system of Pomeranchuk
stability criteria \cite{Peth87} for isoscalar and isovector motions~:
\begin{eqnarray}
& & 1 + F_0  >  0~,                                   \label{Pom1} \\ 
& & 1 + F_0' >  0~,                                   \label{Pom2}
\end{eqnarray}
where $F_0 \equiv F_0^{nn} + F_0^{np}$ and 
$F_0' \equiv F_0^{nn} - F_0^{np}$ are the isoscalar and the isovector Landau
parameters \cite{Migdal}.

\newpage

\section*{ Figure captions }

\begin{description}

\item[Fig. 1] Spinodal boundaries in density-temperature plane 
at different asymmetries (a) and in density-asymmetry plane
at different temperatures (b). Instability regions are under
curves.

\item[Fig. 2] Growth rate of instability as a function of the 
wave vector, as calculated from the dispersion relation (\ref{drel})
for $\rho = 0.2\rho_0$, $T = 5$ MeV (short-dashed lines);
$\rho = 0.4\rho_0$, $T = 5$ MeV (solid lines); 
and $\rho = 0.4\rho_0$, $T = 10$ MeV (long-dashed lines).
Lines are labeled with the asymmetry value $I$. 

\item[Fig. 3] Wave length $\lambda_0$ (a) and growth time $t_0$ (b)
of the most unstable mode as a function of the asymmetry $I$ for 
$\rho^{(0)} = 0.2\rho_0$ (dashed lines) and $\rho^{(0)} = 0.4\rho_0$ 
(solid lines). Lines are labeled with the temperature $T$ in MeV.

\item[Fig. 4] Perturbation asymmetry $I_{pt}$ versus initial
asymmetry $I$ for different initial densities and temperatures :
$\rho^{(0)} = 0.4\rho_0$ and $T = 5$ MeV (solid line), 
$\rho^{(0)} = 0.2\rho_0$ and $T = 5$ MeV (short-dashed line), 
$\rho^{(0)} = 0.2\rho_0$ and $T = 10$ MeV (long-dashed line).

\item[Fig. 5] Time evolution of the density $\rho(x,y)$ in the plane
$z=0$ as given by the test particle code for initial density 
$\rho^{(0)}=0.4\rho_0$, at temperature $T=5$ MeV and asymmetries $I=0$ 
(a) and $I=0.5$ (b). Upper panels show contour plots of the function
$\rho(x,y)$ and lower panels report the corresponding two-dimensional 
surfaces. The density is in units of $fm^{-3} $ x 10$^3$. 

\item[Fig. 6] Time dependence of the density variance 
(a) and of perturbation asymmetry (b) (see text for definitions)
in test particle simulations at initial density 
$\rho^{(0)}=0.4\rho_0$ and temperature $T=5$ MeV for initial
asymmetries $I=0$, $0.25$ and $0.5$ (solid, long- and short-dashed
lines respectively). The straight lines 
on Fig. 6a show linear fits to the initial stage of the SD. 

\item[Fig. 7] Time evolution of neutron (thick solid lines) and 
proton (thin solid lines) abundances (a) and asymmetry (b) in different 
density bins. Calculations refer to the case with initial temperature 
$T=5$ MeV. Initial values of density $\rho^{(0)} = 0.4\rho_0$ and asymmetry 
$I=0.5$ are indicated by dashed lines.

\item[Fig. 8] Time evolution of asymmetries in liquid (solid lines) and in 
gas (long-dashed lines) for initial asymmetries (shown by short-dashed 
lines) $I=0$, $0.25$ (a) and $I=0$, $0.5$ (b). 

\item[Fig. 9] Ratio of total number of particles in gas to total number of 
particles in liquid (a) and number of neutrons and protons in gas (b) 
versus time.

\item[Fig. 10] Asymmetries in gas (squares) and in liquid (circles)
as functions of the initial asymmetry. Long-dashed (solid) lines correspond 
to the "{\it freeze-out}" time $400~(150)$ fm/c. 
In the liquid phase there is no 
appreciable difference between the two lines.

\end{description}


\begin{thebibliography}{99}
\bibitem{Comm0} The most general expression (see Eq.(5) in Ref.\cite{Haensel})
for the quasiparticle interaction amplitude should include also a term, 
caused by the Coulomb interaction, $\propto(\tau_{q_1} - \tau_{q_2})$, 
which does not conserve the total isospin of the colliding quasiparticles.
In our work we drop this term for simplicity.
\bibitem{Comm} These parts become real phases at the late stage of
the "$SD$" ($t > 200~fm/c$). However below, for brevity, we
call liquid- (gas-) like part as liquid (gas) during all the
evolution.
\end{thebibliography}
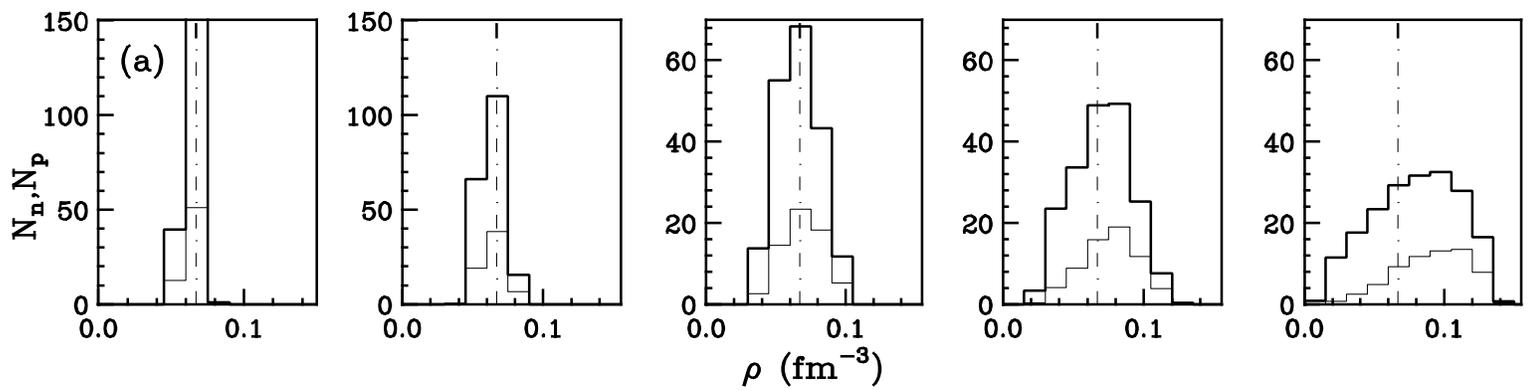
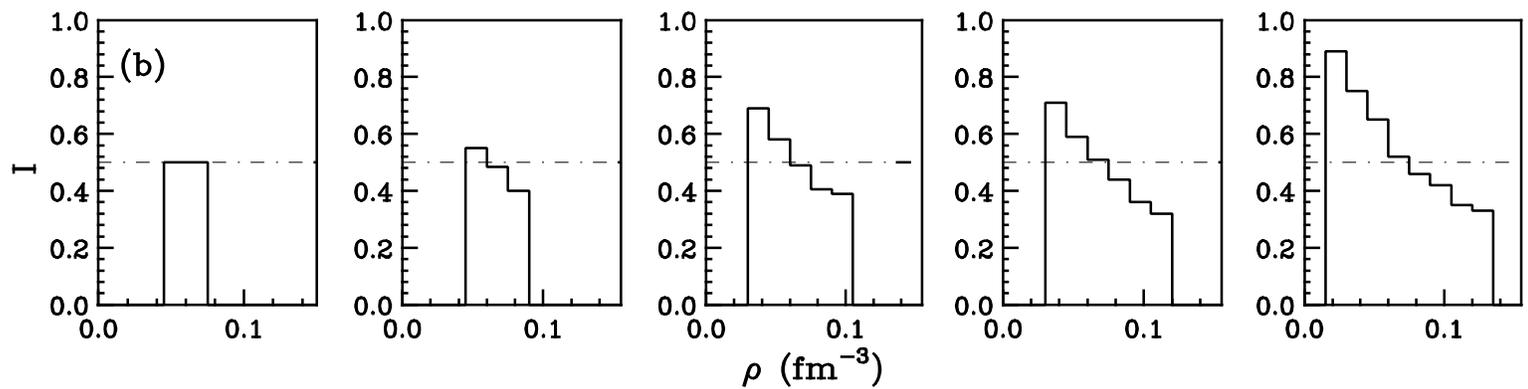
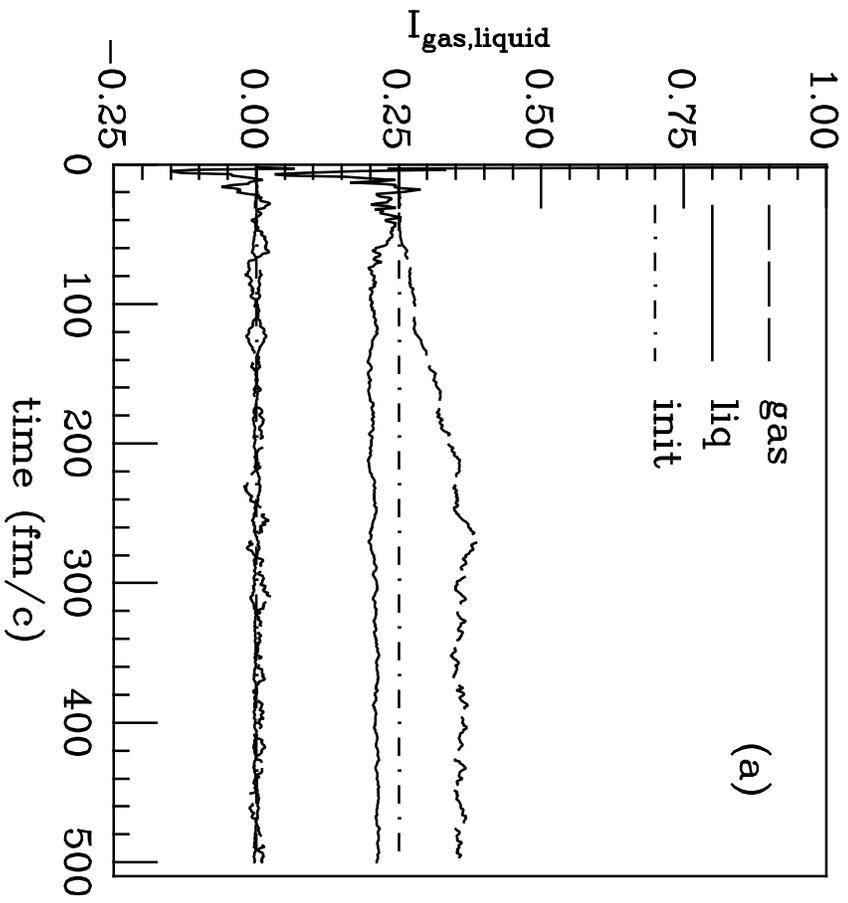
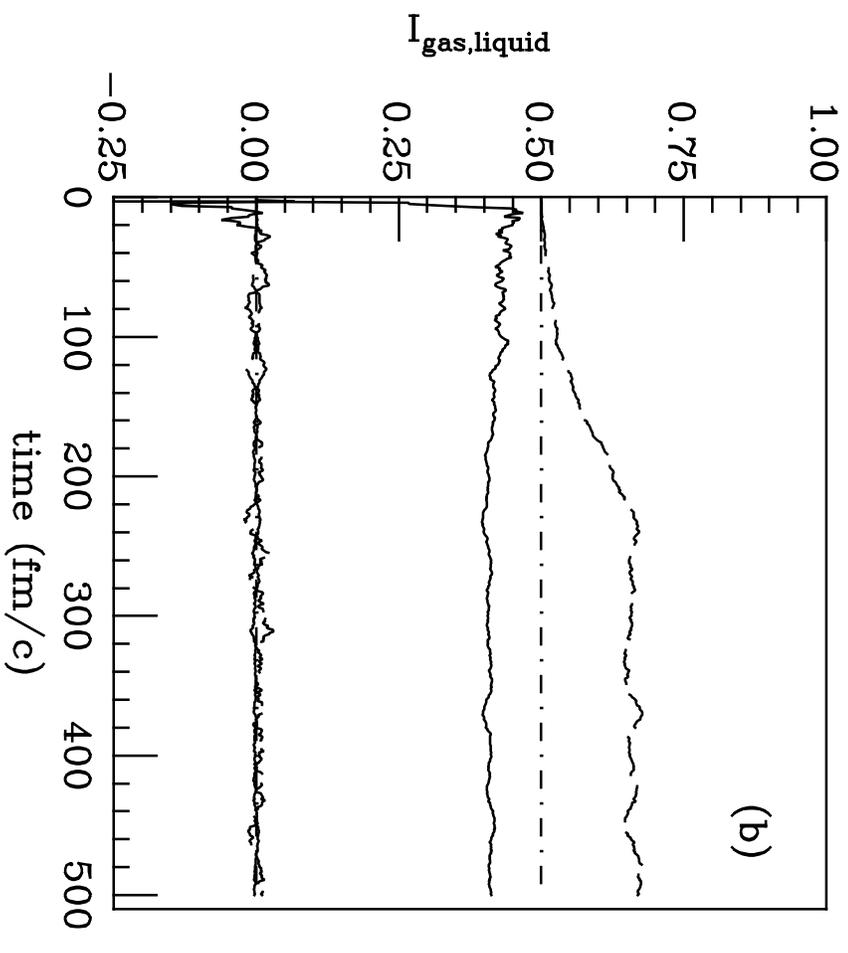
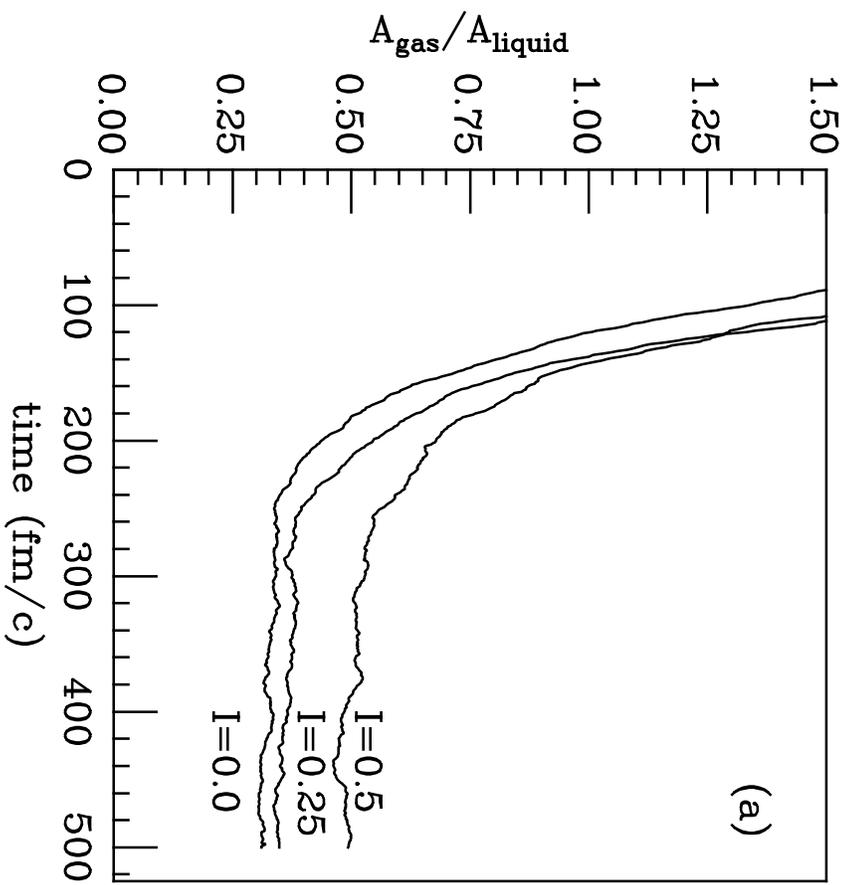
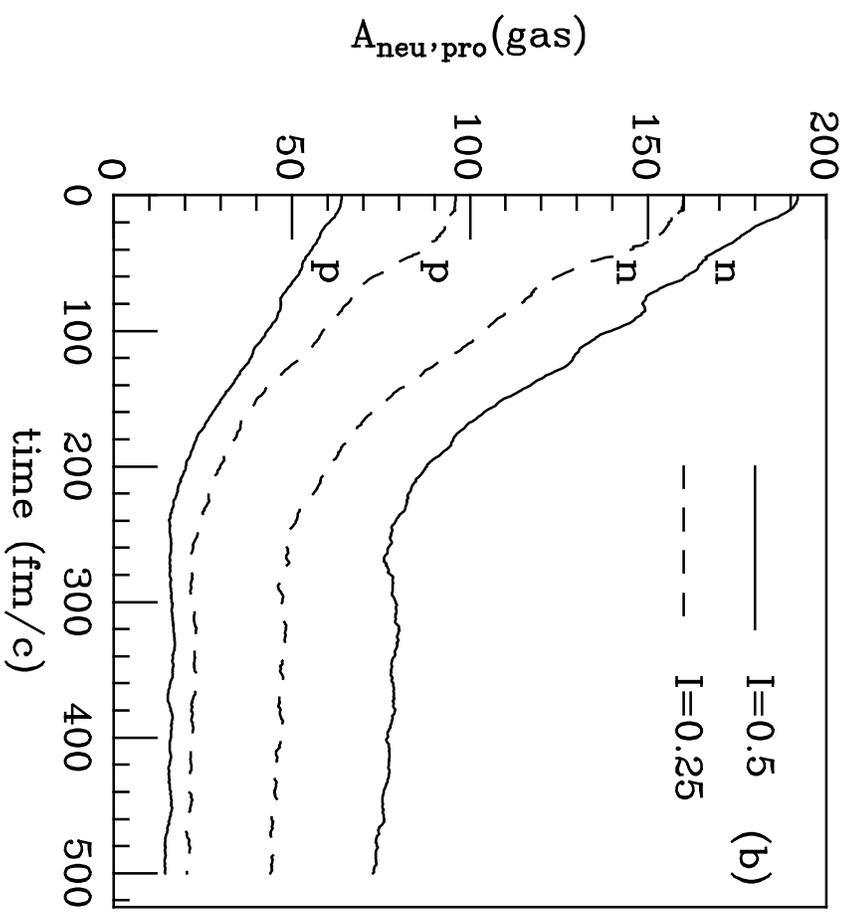
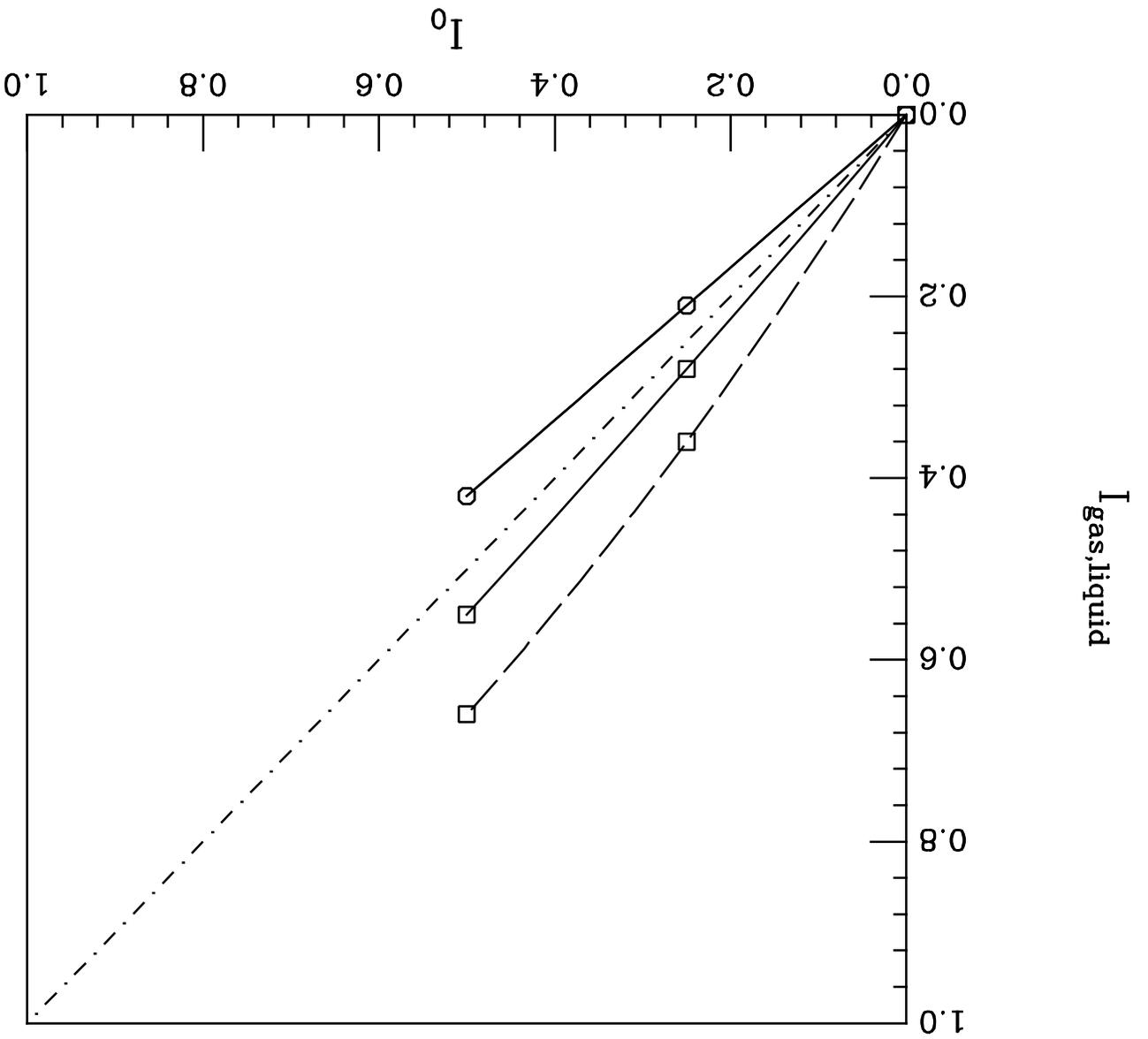
\end{document}